\documentclass[aps,prl,amsmath,amssymb,reprint,groupedaddress]{revtex4-2}

\usepackage{graphicx}
\usepackage[dvipdfmx]{color}
\usepackage{dcolumn}
\usepackage{bm}
\usepackage[normalem]{ulem}

\newcommand{\txtr}[1]{\textcolor{red}{#1}}

\newcommand{\txtb}[1]{\textcolor{blue}{#1}}

\begin{document}
\title{Flux roughening in spin ice with mixed $\pm J$ interactions}
\author{Masaki Kato$^1$, Hiroyasu Matsuura$^1$, Masafumi Udagawa$^2$ and Masao Ogata$^{1,3}$}
\affiliation{$^1$Department of Physics, University of Tokyo, Hongo, Bunkyo-ku, Tokyo 113-0033, Japan \\
$^2$Department of Physics, Gakushuin University, Mejiro, Toshima-ku, Tokyo 171-8588, Japan \\
$^3$Trans-Scale Quantum Science Institute, University of Tokyo, Bunkyo, Tokyo 113-0033, Japan}

\date{\today}

\begin{abstract}
Spin ice presents a typical example of classical spin liquid, where conserved magnetic fluxes emerge from microscopic spin degrees of freedom.
In this letter, we investigate the effect of perturbation by magnetic charge disorder in two-dimensional spin ice. 
To this aim, we develop a novel cluster update algorithm, which enables fast relaxation of magnetic charges.
The efficient Monte Carlo calculation reveals a drastic change of spin structure factor as doping magnetic charges: the pinch point, characterizing the spin ice, is gradually replaced by a diffusive peak.
We derive an analytical relation connecting the flux fluctuation and the spin structure factor, and explain the evolution of diffusive peak in terms of the roughening of magnetic fluxes.
\end{abstract}

\maketitle

{\it Introduction}.---
Recently, quantum and classical spin liquids attract considerable attention due to their exotic properties and potential application to information devices~\cite{pachos2012introduction,Balents:2010aa,Knolle:2019aa}.
The unusual properties of spin liquids are attributed to the nonlocal nature of these systems: global objects, such as fluxes and vortices emerge from microscopic spin degrees of freedom~\cite{castelnovo2008magnetic,PhysRevB.69.064404,PhysRevB.70.094430,PhysRevLett.86.268}. They act as a main carrier of energy and information to produce many unusual features of the system~\cite{PhysRevLett.122.117201,petrova2015hydrogenic,wan2016spinon,PhysRevLett.120.167202,kourtis2016free,Yan:2021tm,PhysRevB.96.195127,PhysRevB.96.085136,PhysRevB.101.064428,PhysRevB.100.014417,Pan:2016ul,Tokiwa:2016wi,PhysRevB.71.224109,ralko2007crystallization,udagawa2002exact,PhysRevB.68.064411,Jaubert:2009wr}.

Spin ice gives a typical example of classical spin liquid, which now has a long history of intensive research~\cite{udagawa2021spin}. In spin ice, spins satisfy a strict local constraint of 2-in 2-out ice rule, which can be mapped to a global magnetic fluxes under a Gauss' law.
This flux picture provides a systematic description on many aspects of spin ice, such as the origin of singularity in the magnetic structure factor known as a pinch point~\cite{PhysRevLett.93.167204,fennell2009magnetic}.
Moreover, one can associate a winding number with the flux itself, and make a topological classification of degenerate ground states.
The notion of flux is also useful to the computational and memory device application of artificial spin ice~\cite{Wang:2006uw,Skjaervo:2020wy,schiffer2021artificial}.

Given these crucial roles of the magnetic fluxes, its response to external perturbation is of natural interest.
In general, the information stored in nonlocal objects is robust against local perturbations.
On one hand, this robustness is favorable for the purpose of e.g. memory storage, since the stored information is hard to dissipate.
On the other hand, the robustness results in the difficulty for manipulation, since one usually uses local probes to control information. These conflicting aspects require us to sophisticate our knowledge: it is crucial to classify the response of magnetic fluxes, according to the types of external perturbations.

One remarkable case study in this direction is the introduction of non-magnetic impurities or spin vacancies~\cite{snyder2002dirty,kajvnakova2004thermodynamic,PhysRevB.73.174429,PhysRevLett.114.247207,PhysRevB.92.085144,PhysRevE.97.042132,PhysRevE.99.022138}.
Vacancies unexpectedly introduce emergent dipoles in spin ice as a result of fractionalization, which serves as an effective low-energy degrees of freedom, leading to a topological glassy state~\cite{PhysRevLett.114.247207}.
In this letter, we examine another possibility of perturbation by magnetic charge disorder induced by ferromagnetic interactions.
This setting may be relevant to the application of spin ice to memory device for computation, since the induced magnetic charges may be used as elemental bits, storing binary information as the sign of charges.
The system with mixed interaction signs may also be relevant in considering a class of pyrochlore oxides, such as iridates~\cite{matsuhira2002low,PhysRevLett.108.066406,PhysRevLett.111.036602,nakatsuji2006metallic,PhysRevLett.98.057203,PhysRevB.92.054432,PhysRevLett.115.056402,PhysRevB.89.075127,matsuhira2013giant,tomiyasu2012emergence,uehara2022phonon,doi:10.1142/S2010324715400044}, 
where the interaction between rare-earth moments are in a subtle competition between ferromagnetic and antiferromagnetic~\cite{matsuhira2002low}. 

In this letter, we consider the two-dimensional spin ice model, i.e. the nearest-neighbor antiferromagnetic Ising model on a checkerboard lattice with a certain fraction of ferromagnetic plaquettes.
The magnetic charges induced on ferromagnetic plaquettes cause a severe difficulty of slow relaxation.
To overcome this problem, we develop a novel algorithm, which we name Zero-energy Cluster (ZEC) update.
This new algorithm enables us to explore the gradual destruction of spin ice by the induced magnetic charges.
We find that the collapse of spin ice is characterized by the evolution of magnetic structure factor: the iconic pinch point of spin ice is replaced by a diffusive peak.
We derive an analytical relation between the magnetic flux fluctuation and the structure factor at a special wavenumber, and find that a flux roughening causes this novel instability of spin ice.

{\it Model}.---
We consider the nearest-neighbor Ising model on a checkerboard lattice of $N_{\rm P}\equiv N\times N$ plaquettes (squares with diagonals) with periodic boundary conditions. Ising spins $\sigma_j=\pm1$ are placed on all $N_{\rm{site}}\equiv2N_{\rm P}$ lattice points [Fig.~\ref{fig:Fig1}(a)]. If we consider plaquette-dependent interactions, the Hamiltonian reads
\begin{eqnarray}
\mathcal{H} = \sum_{\langle i,j\rangle}J_{ij}\sigma_i\sigma_j = \frac{1}{2}\sum_{p}J_pQ_p^2 + {\rm Const.}.
\label{eq:Eq1}
\end{eqnarray}
On each plaquette $p$, we have introduced a magnetic charge $Q_p\equiv \eta_p\sum_{i\in p}\sigma_i$. The sign factor $\eta_p$ takes $+1$ ($-1$) on the sublattice A (B) of plaquette $p$ [Fig.~\ref{fig:Fig1}(a)], which is essential to make $Q_p$ a conserved charge.

If $J_p>0$, Hamiltonian {(\ref{eq:Eq1}) immediately results in the ice rule $Q_p=0$ at the ground state, which leaves macroscopic degeneracy~\cite{ramirez1999zero,PhysRevLett.18.692}.
In this work, we choose $J_p$ randomly according to the binary distribution $P(J_p)$:
\begin{eqnarray}
J_p = \begin{cases} 
-J & P(-J)=p_{\rm{F}},\\
J  & P(+J)=1-p_{\rm{F}}.
\end{cases}\
\label{eq:Eq2}
\end{eqnarray}
$p_{\rm{F}}$ means a fraction of ferromagnetic plaquettes: $p_{\rm{F}}=0$ ($1$) corresponds to the checkerboard spin ice (pure ferromagnetic Ising model).

The ferromagnetic plaquettes induce finite magnetic charges at low temperatures. At least for a small number of ferromagnetic plaquettes, the ground state keeps macroscopic degeneracy, satisfying a modified ice rule: $Q_p=0\ (\pm 4)$ for antiferromagnetic (ferromagnetic) plaquettes. A convenient illustrative method to describe this modified ice rule is the arrow representation: replace a spin $\sigma_j=+1 $ ($-1$) with an arrow pointing from the plaquette of A (B) sublattice to B (A) sublattice [Fig.~\ref{fig:Fig1}(b)]. In this expression, each ground state corresponds to a configuration of six-vertex model with all-in/all-out vertices of ferromagnetic plaquettes. This expression is also suitable for the formulation of columnar transfer matrix to calculate the observables at the ground state.

Another advantage of the arrow representation is that we can visualize a conserved magnetic flux [Fig.~\ref{fig:Fig1}(b)]. To see this, let us draw a line parallel to the $y$-axis between two adjacent columns of plaquettes. We define $W^x_{j_x}$ as the difference between the number of right-directed arrows and that of left-directed arrows on the line. The magnetic flux is conserved in a sense that $W^x_{j_x}$ does not depend on the position of the line as long as $Q_p=0$ is satisfied. More precisely, the magnetic charge $Q_p$ serves as a source of the flux: When a finite charge $Q_p$ exists, the value of $W^x_{j_x}$ changes by $Q_p$. In this sense, $W^x_{j_x}$ deserves the name of ``magnetic flux". 
We will come back to a formal definition of magnetic flux and its relation to an observable quantity later.

\begin{figure}[h]
\begin{centering}
\includegraphics[width=\linewidth]{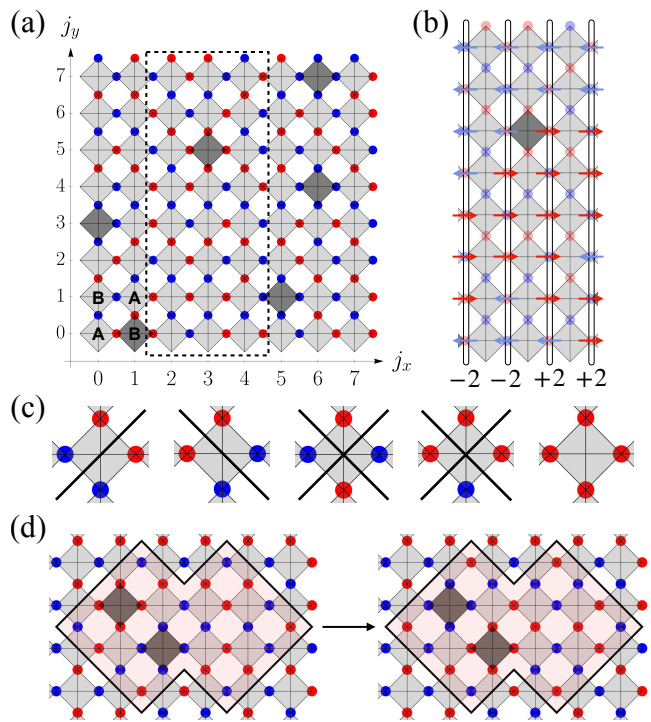}
\caption{(a) The lattice geometry is shown for $N=8$.
A checkerboard lattice is placed on the $x$-$y$ plane, with the origin at the left-bottom plaquette.
The space between neighboring plaquettes is set to be $1$: The plaquettes are located at ${\bm{R}}_p=(j_x, j_y)$, and the spins are located at ${\bm{r}}_j=(j_x, j_y+\frac{1}{2})$ and $(j_x+\frac{1}{2}, j_y)$ with integers, $0\leq j_x, j_y\leq N-1$.
One of the ground states of Hamiltonian (\ref{eq:Eq1}) is shown here. The red (blue) circles represent $\sigma_j=+1$ ($-1$). The light (dark) grey plaquettes have antiferromagnetic (ferromagnetic) interactions: $J_p=+1$ ($-1$). The plaquettes are classified into two sublattices ``A" and ``B". (b) Arrow representation for the spin configuration in a dashed box in (a). Right(left)-directed arrows are depicted in red (blue).
The numbers at the bottom show $W_{j_x}^x =$ (the number of \txtr{$\rightarrow$}) - (the number of \txtb{$\leftarrow$}) on a line separating two plaquette columns at $x=j_x$ and $j_x+1$. 
On the both sides of the central column, $W_{j_x}^x$ changes by $4$, due to the ferromagnetic plaquette supporting a charge $Q_p=4$.
(c) Examples of bonds drawn in the ZEC update. (d) Example of a closed region separated by bonds in the ZEC update.}
\label{fig:Fig1}
\end{centering}
\end{figure}

{\it Method}.---
We study the low-temperature property of Hamiltonian (\ref{eq:Eq1}) by Monte Carlo method. 
In the simulation of spin-ice-type models, the loop update algorithm is usually available to accelerate relaxation at low temperatures~\cite{barkema1998monte,melko2001long,PhysRevLett.100.067207}. 
This update method, however, does not work for the relaxation of magnetic charges.
The basic idea of the loop update is to find a tensionless loop by Brownian motion, which premises the ice rule and is effective only in the spin ice manifold.
A variant of the loop update was also proposed to transfer a charge along a string connecting positive and negative charges~\cite{PhysRevLett.119.077207}. 
However, this algorithm is also ineffective for the present purpose, since the stable charge values of ferromagnetic plaquette are well separated, $Q_p=\pm4$.
With this charge transfer algorithm, we need to transfer $\Delta Q=8$ by finding four strings connecting a charge pair, which inevitably make the program code extremely cumbersome.

Considering these difficulties in the string-based algorithms, we turn our attention to a cluster rather than a string, and propose Zero-energy Cluster (ZEC) update as introduced below.
The basic idea of the ZEC update is to find a cluster of spins flippable without any energy cost.
By repeating cluster flips that contain different sets of magnetic charges, one can relax charges efficiently.
The protocol of ZEC update is simply summarized as:
\noindent
\begin{itemize}
\item[(i)] Draw a bond on each plaquette, if the interactions across the bond sum up to zero [Fig.~\ref{fig:Fig1}(c)].
\item[(ii)] Choose a closed region surrounded by the bonds and flip all the spins inside the boundary [Fig.~\ref{fig:Fig1}(d)].
\end{itemize}
In step (i), we must select the bonds on only one of the two sublattices of the dual square lattice
Otherwise, the detailed balance condition is violated.
For details of this point and several practical information, such as the procedure to identify the boundaries, see Supplemental Material~\footnote{See Sec. A-C in Supplemental Material.}.
In the following, we use the ZEC update in parallel with the single spin update based on the heat bath algorithm and the loop update.

\begin{figure}[h]
\begin{centering}
\includegraphics[width=0.9\linewidth]{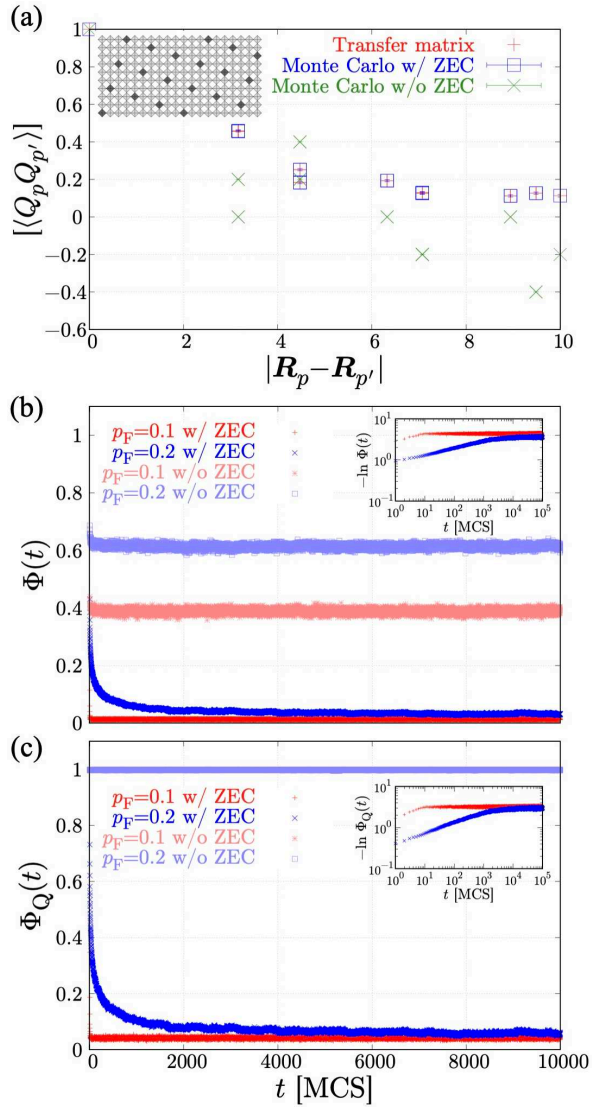}
\caption{(a) Comparisons of the charge correlation $[\langle Q_pQ_{p'}\rangle]$ on ferromagnetic plaquettes $p, p'$ between the exact transfer matrix method and the Monte Carlo simulation with and without the ZEC update, for the periodic array of ferromagnetic plaquettes embedded in the checkerboard lattice of $20\times10$ plaquettes, as shown in the inset. (b) $\Phi(t)$ and (c) $\Phi_{\rm{Q}}(t)$ for $p_{\rm{F}}=0.1$ and $0.2$ with and without the ZEC update. The insets show (b) $-\ln\Phi(t)$ and (c) $-\ln\Phi_{\rm{Q}}(t)$.}
\label{fig:Fig2}
\end{centering}
\end{figure}

{\it Benchmark}.---
To justify our algorithm, we show the comparison with exact transfer matrix calculation for a small system of $20\times 10$ plaquettes with a periodic array of ferromagnetic plaquettes [inset of Fig.~\ref{fig:Fig2}(a)]. Figure \ref{fig:Fig2}(a) shows the charge correlation on ferromagnetic plaquettes $[\langle Q_pQ_{p'}\rangle]$ for different three methods: transfer matrix (red), Monte Carlo method with the ZEC update (blue) and Monte Carlo method without the ZEC update (green).
The conventional algorithm leads to substantial deviation even for this size of small cluster. With the ZEC update, the difference from the exact result is within the error bar of Monte Carlo sampling.
We have made comparisons for a variety of periodic charge patterns and found that the ZEC update gives accurate results up to $p_{\rm{F}}\ \lesssim\ 0.2$. 
The ZEC update is still useful above this upper bound to some extent. Yet certain local arrangements of ferromagnetic plaquettes interrupt the performance of this algorithm. For this problem, see Supplemental Material~\footnote{See Sec. D in Supplemental Material.}.

To demonstrate the efficiency of the ZEC update for fast relaxation, we compute the spin and charge autocorrelation functions defined as, $\Phi(t)\equiv|[\frac{1}{N_{\rm{site}}}\sum_i\langle\sigma_i(0)\sigma_i(t)\rangle]|$ and $\Phi_{\rm{Q}}(t)\equiv|[\frac{1}{N_{\rm F}}\sum_{p\in\boxtimes_{\rm{F}}}\langle Q_p(0)Q_p(t)\rangle]|$ respectively, with $t$ being the Monte Carlo steps after the equilibration.
Here, $\langle\cdots\rangle$ and $[\cdots]$ mean the thermal average and the random average over the configurations of ferromagnetic plaquettes respectively.
Note that the charge correlation is defined for the ferromagnetic plaquettes.
Figs.~\ref{fig:Fig2}(b) and (c) show $\Phi(t)$ and $\Phi_{\rm{Q}}(t)$ calculated for $N_{\rm P}=10^4$ and $p_{\rm{F}}=0.1$ and $0.2$ at $T/J=10^{-2}$, with and without the ZEC update. 
Fig.~\ref{fig:Fig2}(b) clearly shows that the ZEC update improves the spin relaxation considerably. With only single and loop updates, $\Phi(t)$ stays $\sim 0.4$ ($0.6$) for $p_{\rm{F}}=0.1$ ($0.2$) after a long time. With the ZEC update, $\Phi(t)$ quickly falls off to the correct value near $\sim0$. This tendency is clearer in the charge relaxation, as shown in Fig.~\ref{fig:Fig2}(c). Without the ZEC update, the charge autocorrelation does not show the slightest sign of relaxation and keeps the value $\sim1$. This persistence of magnetic charge is quickly improved by the introduction of ZEC update.

We also note on the characteristic relaxation process in the ZEC update algorithm.
In the insets of Figs.~\ref{fig:Fig2}(b) and (c), we show the logarithmic plots of $-\ln\Phi(t)$ and $-\ln\Phi_{\rm{Q}}(t)$, respectively.
Towards the equilibrium values, they show nearly linear behavior, i.e. the relaxation process can be fitted by a stretched exponential, $\sim\exp\left(-(t/\tau)^{\beta}\right)$ with $\beta$ dependent on the value of $p_{\rm{F}}$.
The stretched exponential relaxation is often discussed in glasses and highly frustrated systems~\cite{de1985stretched,johnston2005dynamics}, where complex hierarchy of energy scales is suspected. 
The cluster sizes flipped through the ZEC update also makes a complex distribution, which may underlie this nontrivial relaxation dynamics.

\begin{figure*}[t]
\begin{centering}
\includegraphics[width=\linewidth]{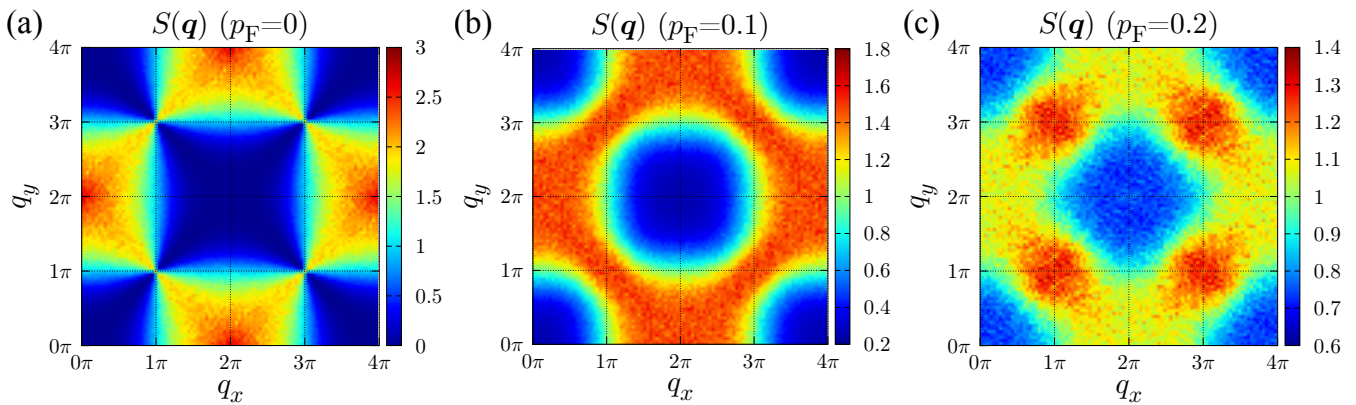}
\caption{The magnetic structure factor $S(\bm{q})$ calculated by Monte Carlo simulation for $N_{\rm P}=10^4$ for (a) $p_{\rm{F}}=0$, (b) $p_{\rm{F}}=0.1$ and (c) $p_{\rm{F}}=0.2$.}
\label{fig:Fig3}
\end{centering}
\end{figure*}

{\it Structure factor}.---
We now investigate the physical properties of the system using the ZEC update algorithm.
As a physical observable, we focus on the spin structure factor, $S(\bm{q})=[\langle\sigma_{-{\bm{q}}}\sigma_{\bm{q}}\rangle]$, where $\sigma_{\bm{q}}\equiv\frac{1}{\sqrt{N_{\rm{site}}}}\sum_{j}\sigma_j e^{i\bm{q}\cdot\bm{r}_j}$.
Figure~\ref{fig:Fig3} shows $S(\bm{q})$ obtained by the Monte Carlo simulation for $N_{\rm p}=10^4$ for different fractions of ferromagnetic plaquettes at the temperature, $T/J=10^{-2}$.
At $p_{\rm F}=0$, all the plaquettes are antiferromagnetic, and the system forms a perfect spin ice at zero temperature.
Reflecting the spin ice formation, $S(\bm{q})$ exhibits a singularity at $\bm{q}=\bm{q}_{\rm p}\equiv(\pi, \pi)$ and equivalent wavenumbers [Fig.~\ref{fig:Fig3}(a)].
This is the pinch point, an iconic character of spin ice, reflecting the divergence-free nature of magnetic fluxes~\cite{fennell2009magnetic,PhysRevLett.93.167204}.

As the number of ferromagnetic plaquettes increases, the ice rule is gradually modified.
As shown in Fig.~\ref{fig:Fig3}(b), the pinch point quickly blurs and is replaced by a ridge-like structure, which is reminiscent of the neutron scattering images for several spin ice candidate materials~\cite{Kimura:2013vy,Petit:2016ti,PhysRevB.94.165153,castelnovo2019rod}. 
As the fraction of ferromagnetic plaquettes is further increased [Fig.~\ref{fig:Fig3}(c)], even a broad peak develops at $\bm{q}_{\rm p}$ where the pinch point was originally placed.
In general, the destruction of the pinch point reflects the instability of spin ice, and the structural change of $S(\bm{q})$ around ${\bm{q}}_{\rm p}$ is used to classify the type of instability~\cite{PhysRevLett.110.107202,castelnovo2019rod,PhysRevLett.119.077207,Rau:2016aa,PhysRevB.94.104416,PhysRevB.98.144446}.
However, to our knowledge, the growth of a peak at $\bm{q}_{\rm p}$ has rarely been reported either in theories or in experiments on classical spin ice, with a possible exception in a three-dimensional quantum spin ice candidate system~\cite{petit2012spin}.

{\it Magnetic fluxes}.---
Actually, the value of $S(\bm{q}_{\rm p})$ carries important information on the nonlocal property of the system as discussed below.
This wavenumber, $\bm{q}_{\rm p}=(\pi, \pi)$ has already drawn attention as the place of the singular pinch point. However, here we would like to focus on the value of $S(\bm{q}_{\rm p})$ itself.
Indeed, even after the spin ice is destroyed and the singularity is washed away, $S(\bm{q}_{\rm p})$ stores important information on the global fluxes of the system. This relation also gives a precious example in which nonlocal character of the system can be accessed through local observables~\cite{PhysRevX.3.011014,PhysRevB.102.214427}.

To clarify the connection between the magnetic fluxes and $S({\bm{q}}_{\rm p})$, we write the formal expression of magnetic flux as
\begin{align}
\begin{cases}
W^x_{j_x} = \sum_{j_y}(-1)^{j_x+j_y}\sigma_{j_x+1/2, j_y},\\
W^y_{j_y} = \sum_{j_x}(-1)^{j_x+j_y}\sigma_{j_x, j_y+1/2}.
\end{cases}
\label{eq:magneticflux}
\end{align}
This expression is equivalent to the intuitive arrow counting presented in Fig.~\ref{fig:Fig1} (b). 
In the expression (\ref{eq:magneticflux}), $(-1)^{j_x+j_y}$ is a sign factor to distinguish A and B sublattices of the plaquettes [Fig.~\ref{fig:Fig1} (a)]. This factor is in turn translated into a phase factor at $\bm{q}_{\rm p}=(\pi, \pi)$. We obtain $\sigma_{\bm{q}_{\rm p}}=\frac{1}{\sqrt{N_{\rm{site}}}}(\sum_{j_x}W^x_{j_x} + \sum_{j_y}W^y_{j_y})$ up to a constant phase, which leads to the relation,
\begin{equation}
S(\bm{q}_{\rm p}) = [\langle (W^x)^2\rangle] = [\langle(W^y)^2\rangle],
\label{eq:Sq_fluxfluctuation}
\end{equation}
where $W^x=\frac{1}{N}\sum_{j_x}W^x_{j_x}$ and $W^y=\frac{1}{N}\sum_{j_y}W^y_{j_y}$. This claims that $S(\bm{q}_{\rm p})$ is proportional to the fluctuation of the global magnetic fluxes.

In terms of Eq.~(\ref{eq:Sq_fluxfluctuation}), it is possible to attribute the evolution of the structure factor to the enhanced fluctuation of the magnetic fluxes.
In the absence of ferromagnetic plaquettes, the flux value is uniform over the system, and the zero-flux configurations have the largest weight. 
Finite-flux configurations occur only occasionally over the Monte Carlo steps.
In the presence of ferromagnetic plaquettes, however, the nature of the flux fluctuation qualitatively changes: The fluctuation becomes ``spatial" rather than ``temporal".
In Fig.~\ref{fig:Fig4}, we show the snapshot of the flux distribution.
As is clearly seen here, the flux $(W^x_{j_x}, W^y_{j_y})$ changes its value at ferromagnetic plaquettes, producing a mosaic pattern.
The accumulated charges $Q_p=\pm4$ on the ferromagnetic plaquettes force the flux value to change by 4.
As the fraction of ferromagnetic plaquettes increases, the flux distribution shows roughening: This enhanced spatial fluctuation of the magnetic fluxes is observed as the diffusive peak of $S(\bm{q})$ at $\bm{q}=\bm{q}_{\rm p}$.

\begin{figure}[h]
\begin{centering}
\includegraphics[width = 1\linewidth]{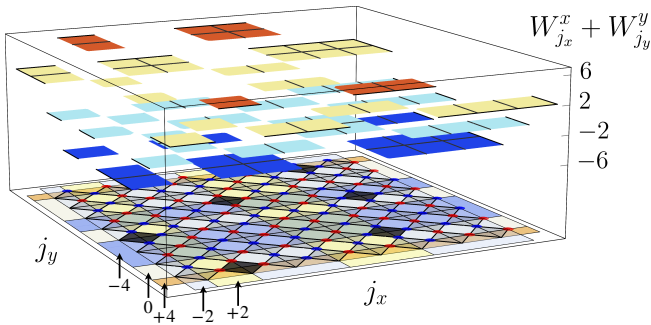}
\caption{Snapshot of the set of the magnetic fluxes $(W^x_{j_x}, W^y_{j_y})$ and their sum $W^x_{j_x}+W^y_{j_y}$. In the presence of charges, the flux varies spatially and its pattern looks like a mosaic exhibiting spatial roughening.}
\label{fig:Fig4}
\end{centering}
\end{figure}

{\it Summary and Discussions}.---
In summary, we have studied the effect of perturbation by magnetic charge disorder in spin ice, by taking the model of checkerboard lattice with mixed $\pm J$ plaquette interactions.
For the relaxation of magnetic charges, we have developed a new type of cluster update algorithm for the Monte Carlo simulation called ZEC.
This numerical innovation enabled us to explore the gradual collapse of spin ice by increasing ferromagnetic plaquettes, which is signaled by the replacement of the pinch point by a diffusive peak.
The evolution of the diffusive peak is attributed to the flux roughening, which is verified on the basis of a firm relation between the flux fluctuation and the spin structure factor at ${\bm{q}}_{\rm p}=(\pi, \pi)$.

The present study can be extended to many directions.
The collapse of pinch point structure has been used to classify the instabilities of spin ice. This theoretical finding must be useful to interpret future experiments.
For example, in pyrochlore iridates, keen competition of ferromagnetic and antiferromagnetic interactions has been actively debated. While the number of available neutron scattering data is limited due to the notoriously high neutron absorption rate of Ir, it is interesting to look for the signature of competing interactions in $S({\bm{q}})$ at finite temperatures.

It is also interesting to proceed our analysis to larger $p_{\rm F}$, where we expect that the flux roughening shows further enhancement. If the density of ferromagnetic plaquettes increases, the system cannot satisfy the modified ice rule any more, and faces severe frustration.
There, an appearance of complex energy landscape is naturally expected, which may give rise to a new type of topological glassy state.

From the technical viewpoint, our ZEC update brings a new powerful tool to analyze systems suffering slow relaxation.
For instance, it is important to apply this algorithm to spin glass models and other type of spin-ice-like systems~\cite{kato2021magnetic,kosaka2018gas,SAMUELSEN19771275,PhysRevB.84.064120,PhysRevResearch.2.043077}, in which slow relaxation always confronts us.
Our new algorithm can be flexibly extended to higher-dimensional systems or to include different degrees of freedom. We would like to leave these fascinating problems for future work.

\begin{acknowledgments}
This work was supported by JSPS KAKENHI (Nos. 20H05655, 20H04463 and 22H01147), MEXT, Japan.
\end{acknowledgments}

\bibliography{checkerboard_prl}

\end{document}


\title{Supplemental Material for ``Flux roughening in spin ice with mixed $\pm J$ interactions''}
\author{Masaki Kato$^1$, Hiroyasu Matsuura$^1$, Masafumi Udagawa$^2$ and Masao Ogata$^{1,3}$}
\affiliation{$^1$Department of Physics, University of Tokyo, Hongo, Bunkyo-ku, Tokyo 113-0033, Japan \\
$^2$Department of Physics, Gakushuin University, Mejiro, Toshima-ku, Tokyo 171-8588, Japan \\
$^3$Trans-Scale Quantum Science Institute, University of Tokyo, Bunkyo, Tokyo 113-0033, Japan}

\date{\today}
\maketitle

\subsection{Summary of the ZEC update algorithm}
\label{sec:summary}
As introduced in the main text, the essence of the Zero-energy Cluster (ZEC) update algorithm is to divide the spins into clusters, which can be flipped without energy cost.
Here, we summarize the algorithm in a more formal way, to make rigorous arguments.

\begin{figure}[h]
\begin{centering}
\includegraphics[width = 6cm]{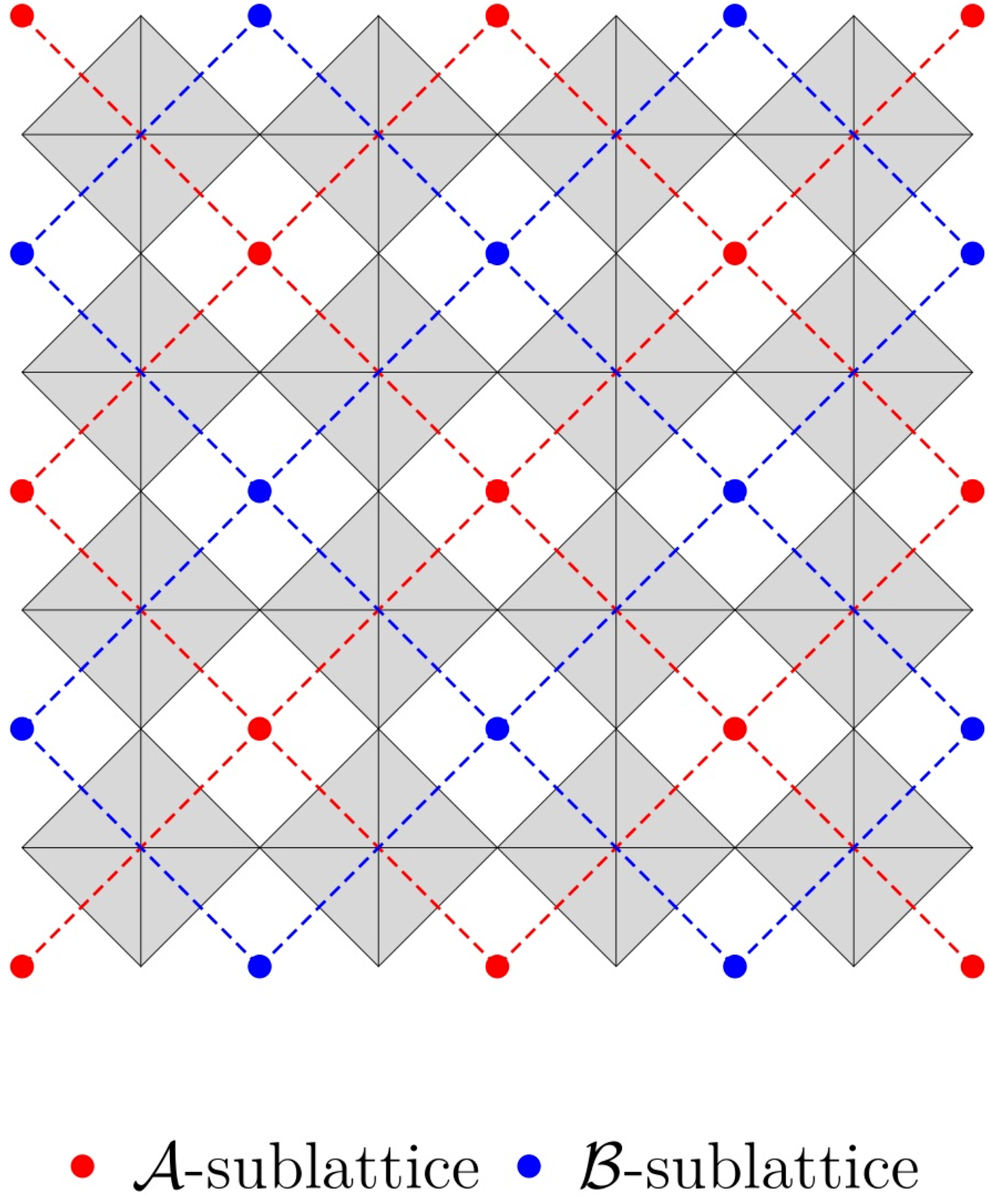}
\caption{Dual sublattices ``$\mathcal{A}$'' and ``$\mathcal{B}$''.}
\label{fig:SupFig1}
\end{centering}
\end{figure}

The checkerboard lattice can be seen as an alternate pattern of plaquettes (with diagonal bonds) and blank squares.
We start with the introduction of dual square lattice by placing sites on blank squares.
We divide the sites on this dual square lattice into two sublattices, $\mathcal{A}$ and $\mathcal{B}$, in a bipartite way [Fig.~\ref{fig:SupFig1}]. 
In the ZEC update algorithm, we connect the neighboring sites on the same sublattice to make a bond across a plaquette, if the interactions across the bond sum up to zero.
We classify the bonds into $\mathcal{A}$-bond and $\mathcal{B}$-bond, according to which sublattice sites the bond connects.
If the bonds make a closed path, we call it a boundary, and classify it into $\mathcal{A}$-boundary and $\mathcal{B}$-boundary in the same way.
We finally define a cluster as a set of spins separated by the boundaries.

Given these definitions, the formal procedure of the ZEC update algorithm can be summarized as below:
\begin{itemize}
\item[(a)] Choose one (dual) sublattice and draw bonds on all the plaquettes
\item[(b)] Allocate boundaries and divide the whole spins into clusters
\item[(c)] Randomly choose one ferromagnetic plaquette
\item[(d)] Flip the spins inside the cluster containing the ferromagnetic plaquette in (c) and return to (a)\\
\end{itemize}

Regarding (a), it is crucial to choose only one sublattice to ensure the detailed balance condition.
We address this point in the next section, Sec.~\ref{sec:DBC}.
To implement the step (b) in a program code, a few practical tips would be helpful, which we introduce in Sec.~\ref{sec:BoundaryConstruction}

\subsection{Detailed balance condition}
\label{sec:DBC}
We show the detailed balance condition is satisfied in the ZEC update algorithm.
In this algorithm, once a cluster is specified and flipped, the flipped cluster can be flipped once again without energy cost.
It naturally defines an inverse process.
To ensure the detailed balance condition, the probabilities of the initial and inverse cluster flips must be the same.

The probability of a cluster flip is equal to the product of two quantities: a probability of choosing one particular cluster and a probability of actually flipping it.
For the latter, we set it to be 1: we adopt the Metropolis rule to flip a cluster with probability 1, once it is chosen. By definition, the cluster flip is an equal-energy transition.
Accordingly, only the former probability is important to satisfy the detailed balance condition: a cluster must be chosen with the same probability in the initial and inverse processes.

To consider the probability of cluster choice, suppose a certain spin configuration with $N_{\rm q}$ ferromagnetic plaquettes with $Q_p=\pm 4$, and divide the spins into zero-energy clusters, where the $i$th cluster has $n_i$ ferromagnetic plaquettes inside. Since our algorithm randomly chooses one ferromagnetic plaquette, and flip the cluster containing it, the probability of choosing the $i$th cluster is $n_i/N_{\rm q}$.
So, if the spins are divided into exactly the same set of clusters after the flip, the inverse process also occurs with the probability $n_i/N_{\rm q}$, and the detailed balance condition is satisfied.

Indeed, to preserve the cluster division after the flip, the sublattice selection plays an important role.
Suppose a cluster flip is made along an $\mathcal{A}$-boundary, then the $\mathcal{A}$-bonds keeps the same configuration after the flip, while the $\mathcal{B}$-bond configuration changes.
To see how the bond configuration changes, let us take one cluster. 
Plaquettes contained in a cluster are divided into two types. (I) \textit{Interior}: all the four spins on a plaquette are inside the cluster, and (II) \textit{surface}: only two spins are inside.
After the cluster flip using an $\mathcal{A}$-boundary, the bonds on the interior plaquettes do not change, while the bonds on the surface plaquettes may change.
More specifically, on a surface plaquette, the original bond, forming a part of the boundary, remains unchanged.
Meanwhile, a new bond may be generated (or an old bond may be erased) perpendicular to the boundary.
This newly generated bond connects the dual plaquettes of $\mathcal{B}$ sublattice, and changes the $\mathcal{B}$-bond configuration.

This clearly points how the detailed balance condition is violated, if $\mathcal{A}$-boundaries and $\mathcal{B}$-boundaries are mixed in a single update: The cluster division changes after the flip, and the flipped cluster is not chosen with the same probability in the inverse process. 
To avoid this difficulty, we only need to use one species of boundaries for one update, then the cluster structure is preserved.
The invariance of the cluster structure may seem unfavorable in terms of ergodicity.
However, if updates with only-$\mathcal{A}$-boundaries and updates with only-$\mathcal{B}$-boundaries are carried out e.g. alternately, one can change cluster divisions and make a search in a large region of phase space without violating the detailed balance condition.

\begin{figure}[h]
\begin{centering}
\includegraphics[width = 8.5cm]{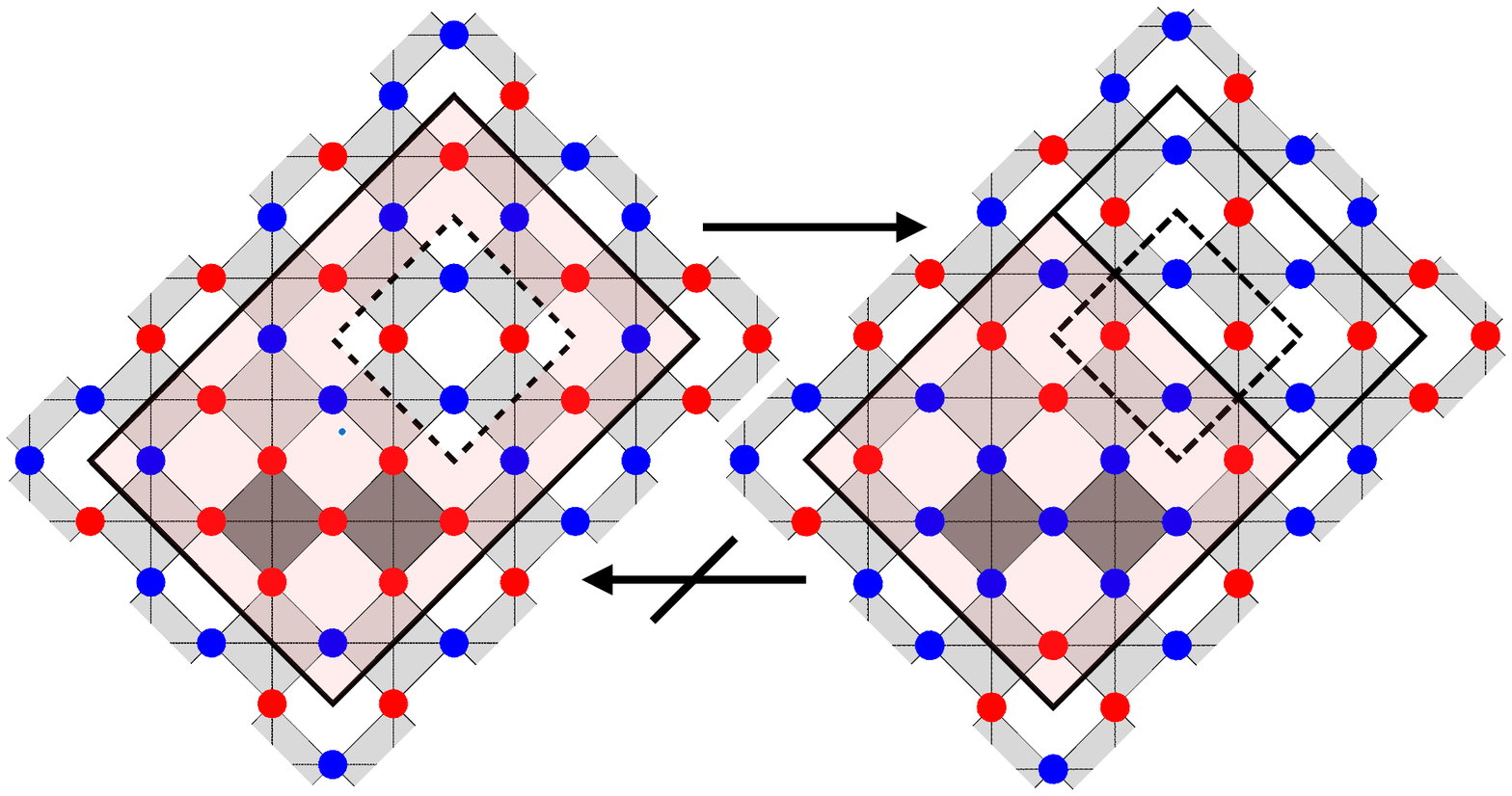}
\caption{In the left figure, the red shaded area is surrounded by the solid line and the dashed line, which belong to different types of boundary. If this region is used as a cluster in an update, the cluster configuration changes to the right and the cluster containing the ferromagnetic plaquettes also changes. Consequently, the detailed balance condition breaks down.}
\label{fig:SupFig3}
\end{centering}
\end{figure}

Figure \ref{fig:SupFig3} shows how the absence of the sublattice selection leads to the breakdown of the detailed balance condition. In this configuration, the region between the solid line and the dashed line (left red shaded area) composes a cluster containing a pair of ferromagnetic plaquettes.
Here, different types of boundaries are used at the same time. After this update, the spin configuration changes from the left to the right in Fig.~\ref{fig:SupFig3}. Then new bonds arise across the original cluster and the cluster containing the ferromagnetic plaquettes shrinks (right red shaded area). This example demonstrates that the mixed use of $\mathcal{A}$- and $\mathcal{B}$-boundaries breaks the detailed balance condition.

\subsection{Details of boundary construction}
\label{sec:BoundaryConstruction}
In this section, we provide practical information to implement the algorithm for numerical calculation.
While we assumed the spins were divided into a certain number of clusters in Sec.~\ref{sec:DBC}, we do not need to identify all the clusters in each update.
Practically, we choose a ferromagnetic plaquette at random and find a cluster that contains this plaquette. 

\begin{figure}[h]
\begin{centering}
\includegraphics[width = 8.5cm]{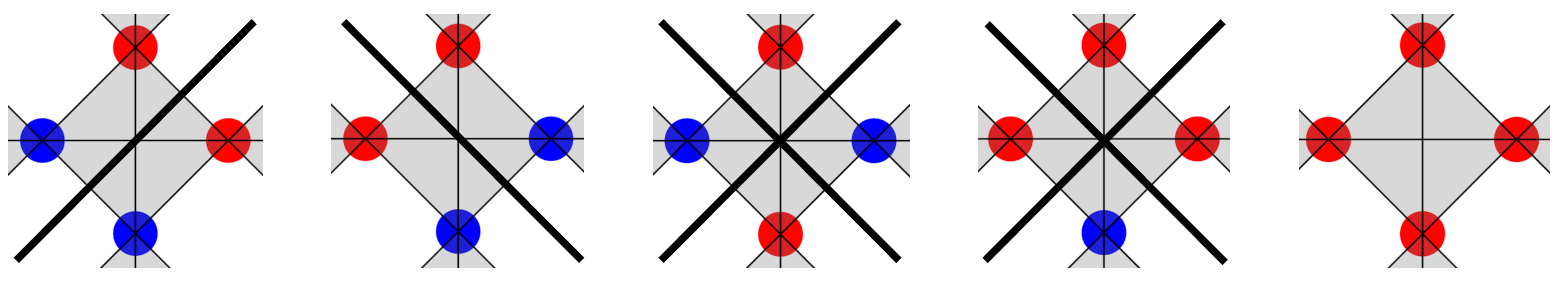}
\caption{Examples of bonds on a plaquette.}
\label{fig:SupFig2}
\end{centering}
\end{figure}

To find a cluster, we resort to a sort of breadth first search.
First, we take one spin on a ferromagnetic plaquette, and move to the neighboring plaquette which shares this spin, and draw bonds, according to the rule in Fig.~\ref{fig:SupFig2}.
Then, we register the connected spins on this neighboring plaquette, i.e. the spins which are not separated by the drawn bonds from the spin on the starting plaquette.
Here, we consider the bonds on only one sublattice and ignore those on the other sublattice, as we have discussed in Sec.~\ref{sec:DBC}.
Then, we repeat this procedure for the four spins on the original plaquette.

Next, we take one of the neighboring plaquettes and choose one of the connected spins.
We then move to the plaquette which shares this spin and draw bonds, according to the rule in Fig.~\ref{fig:SupFig2}, and register the connected spin on this plaquette.
We repeat this procedure to list up all the spins connected to the first plaquette.
This repetition defines a sort of breadth first search algorithm on a plaquette network.
And when this search stops, we can obtain a full list of spins composing the cluster that contains the original ferromagnetic plaquette.

\subsection{Dangerous configurations}
\begin{figure}[h]
\begin{centering}
\includegraphics[width = 4cm]{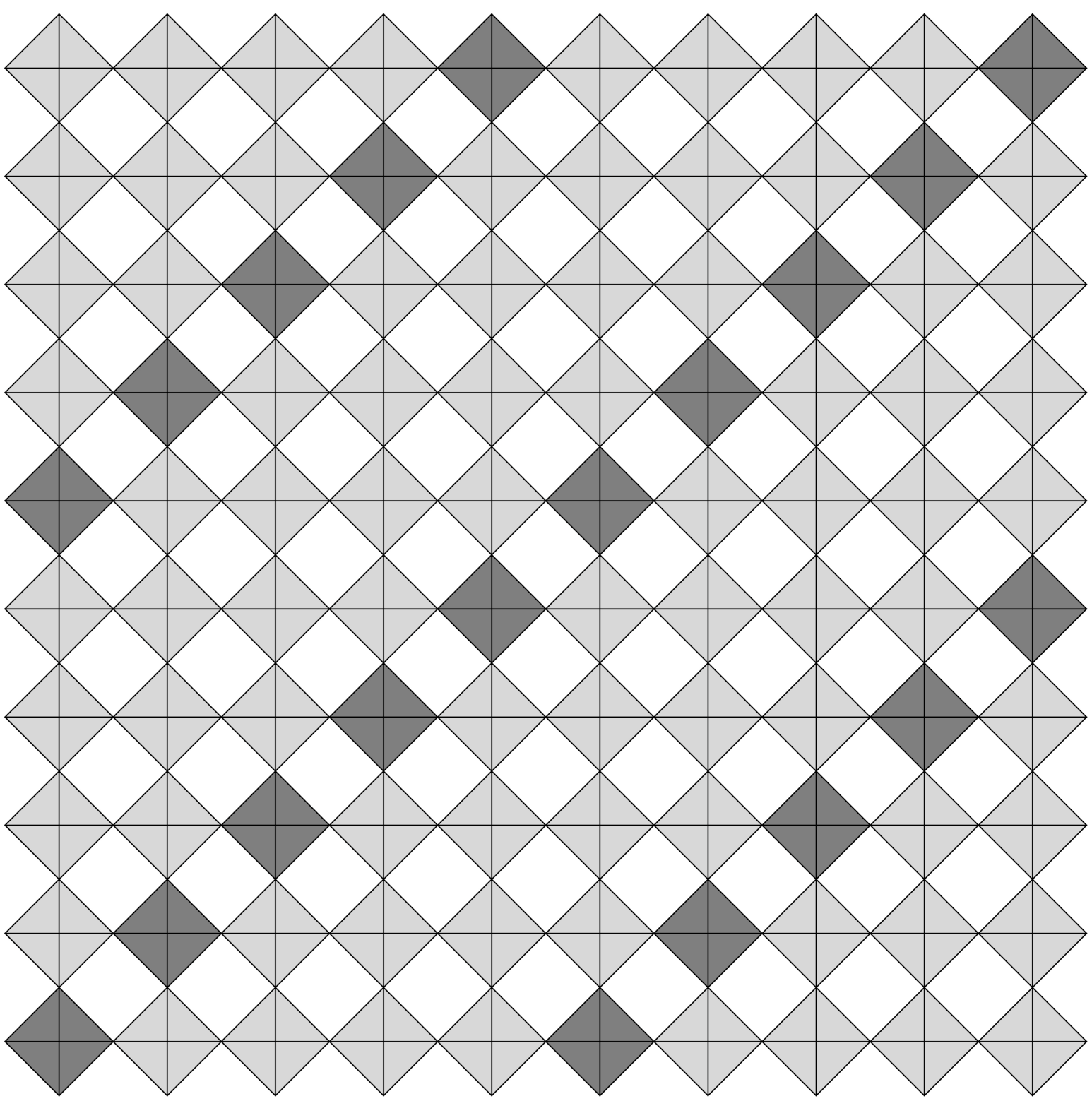}
\caption{Example of a case where the ZEC update does not work with perfect accuracy. Charges on different chains of ferromagnetic plaquettes never form a zero-energy cluster.}
\label{fig:SupFig4}
\end{centering}
\end{figure}

Although the ZEC update does not violate the detailed balance condition, it is not perfect. There are inconvenient layouts of ferromagnetic plaquettes. We show one of such patterns in Fig.~\ref{fig:SupFig4}. In Fig.~\ref{fig:SupFig4}(a), two chains of ferromagnetic plaquettes are embedded in a system of size $10\times 10$. At first glance, it seems that the ZEC update works with complete accuracy since flippable clusters can be easily found. However, we cannot make a cluster that includes plaquettes on both two chains. This indicates that the sum of the charges on each chain is preserved and that the resulting equilibrium depends on the initial charge configuration. 
This kind of unfavorable arrangements will appear more frequently if the density of the ferromagnetic plaquettes increases. 

